\newcommand{\non}{\nonumber}
\newcommand{\R}{{\mathbb R}}   

\def\T{\tiny\mbox{\rm T}}

\def\bfa{\bm a}
\def\bfb{\bm b}

\def\bff{\bm f}

\def\bfu{\bm u}
\def\bfv{\bm v}
\def\bfw{\bm w}
\def\bfx{\bm x}
\def\bfy{\bm y}
\def\bfz{\bm z}
\def\bfA{\bm A}

\def\bfC{\bm C}

\def\bfF{\bm F}
\def\bfG{\bm G}
\def\bfH{\bm H}
\def\bfI{\bm I}

\def\bfK{\bm K}

\def\bfP{\bm P}
\def\bfQ{\bm Q}
\def\bfR{\bm R}

\def\bfU{\bm U}
\def\bfV{\bm V}

\def\bfSigma{\bm\varSigma}

\def\bfxi{\bm\xi}

\def\bfchi{\bm\chi}

\documentclass[letterpaper, 10 pt, conference]{ieeeconf}  

\IEEEoverridecommandlockouts                              

\usepackage{graphics} 
\usepackage{epsfig} 
\usepackage{times} 
\usepackage{amsmath,mathrsfs,amsfonts} 
\usepackage{amssymb}  
\usepackage{bm}
\usepackage{color}
\allowdisplaybreaks[4]
\newtheorem{remark}{Remark}

\title{\LARGE \bf State estimation of a carbon capture process through POD model reduction and neural network approximation}

\author{Siyu Liu$^{1,2}$, Xunyuan Yin$^3$, Jinfeng Liu$^{2}$
\thanks{$^{1}$Siyu Liu is with School of Internet of Things Engineering, Jiangnan University, Wuxi\ 214122, China. 
        {\tt\small syliu12@126.com}}%
\thanks{$^{2}$Jinfeng Liu and Siyu Liu are with Department of Chemical \& Materials Engineering, University of Alberta, Edmonton, AB, Canada, T6G 1H9
    	{\tt\small jinfeng@ualberta.ca}}
    \thanks{$^{3}$Xunyuan Yin is with School of Chemistry, Chemical Engineering and Biotechnology, Nanyang Technological University, 62 Nanyang Drive, Singapore, 637459
    	{\tt\small xunyuan.yin@ntu.edu.sg}}
}

\begin{document}

\maketitle
\thispagestyle{empty}
\pagestyle{empty}


\begin{abstract}

This paper presents an efficient approach for state estimation of post-combustion CO$_2$ capture plants (PCCPs) by using reduced-order neural network models. The method involves extracting lower-dimensional feature vectors from  high-dimensional operational data of the PCCP and constructing a reduced-order process model using proper orthogonal decomposition (POD). Multi-layer perceptron (MLP) neural networks capture the dominant dynamics of the process and train the network parameters with low-dimensional data obtained from open-loop simulations. The proposed POD-MLP model can be used as the basis for estimating the states of PCCPs at a significantly decreased computational cost. For state estimation, a reduced-order extended Kalman filtering (EKF) scheme based on the POD-MLP model is developed. Our simulations demonstrate that the proposed POD-MLP modeling approach reduces computational complexity compared to the POD-only model for nonlinear systems. Additionally, the POD-MLP-EKF algorithm can accurately reconstruct the full state information of PCCPs while significantly improving computational efficiency compared to the EKF based on the original PCCP model.

\end{abstract}

\section{INTRODUCTION}

In recent years, post-combustion CO$_2$ capture plants (PCCPs) have gained significant attention due to their potential in reducing greenhouse gas emissions and mitigate global warming. PCCPs are commonly used in power plants and carbon-intensive industrial processes to separate CO$_2$ from the flue gas \cite{MacDowell2010_RSC}. The operational safety, carbon capture efficiency and economic cost of PCCPs are highly dependent on the performance of the advanced control systems used for regulating the process operations \cite{Manaf2019_JPC}.

Real-time information of the key quality variables of the PCCP is essential for the advanced control system to make the most appropriate decisions for safe and efficient process operation. However, measuring all quality variables online through deploying hardware sensors is unrealistic. Therefore, it is crucial to exploit state estimation to reconstruct the full state information for PCCPs. Unfortunately, results on state estimation of PCCPs have been limited. In \cite{Yin2020_ACSP}, we made an initial attempt on estimating the states of the absorber of a PCCP by developing a distributed moving horizon estimation scheme. However, the other key quality variables associated with the desorption unit and other physical units are not addressed within this framework. Wang et al., presented a robust soft sensor using a neural network and moving horizon estimator to monitor key operating parameters in the carbon capture process \cite{Wang2022_Fuel}. In the context of nonlinear state estimation, there have been some algorithms that have the potential to be leveraged for state estimation of PCCPs, e.g., extended Kalman filtering \cite{DF2022_Auto_LiuSY}, moving horizon estimation methods \cite{Yin2017_Auto}, and particle filter \cite{Arulampalam2002_IEEETSP}. However, most of nonlinear estimation algorithms require accurate first-principles dynamic models of the underlying nonlinear processes. Due to the large scales and complex structures of PCCPs, it can be challenging to conduct first-principles modeling. Additionally, even if a first-principles model is obtained, this type of model that will involve partial differential equations that describe the dynamical behaviors of the absorption and desorption columns will be computationally expensive to simulate, especially considering the cases when optimization-based estimation approaches are used.

Data-driven modeling using neural networks has been widely used as an alternative to first-principles modeling for various nonlinear processes. For example, Jeon et al. utilized the neural networks (NNs) to describe the dynamics of a chemical reactor, which was further used to optimize its control performance \cite{Jeon2022_CCE}. Cheng et al. employed NNs to model the ship motion and improved its navigation accuracy \cite{Cheng2020_IEEEJOE}. Zhao et al., developed reduced-order recurrent neural networks that capture the dominant dynamics of nonlinear systems using an autoencoder \cite{Zhao2022_CERD}. However, these methods may face certain limitations when considering possible applications to PCCPs. Specifically, the complexity of the NN model, including the number of neurons and layers, and the computational cost of training, may increase exponentially with the dimensionality of the input and output data, commonly referred to as the ``curse of dimensionality".


To address these limitations associated with state estimation of PCCPs, we propose a solution that combines data-driven modeling using NNs with model reduction techniques. Proper orthogonal decomposition (POD) has been widely adopted in various engineering fields for reducing the dimensionality of high-dimensional data sets while preserving dominant patterns or features of the data. In the context of control systems, POD can be used to satisfactorily approximate the dynamics of a large-scale nonlinear process in a lower-dimensional state space \cite{Liang2002_JSV,Kerschen2005_ND}. POD holds the promise to lower the dimensionality of the originally high-dimensional state-space model for PCCP. Specifically, our objective is to leverage POD to create a reduced-order machine learning-based model with lower structural complexity. This model can then facilitate reductions in the computational cost for model training and implementation of state estimation based on the reduced-order model.

Motivated by the observations above, we propose a neural network-based state estimation approach for PCCPs using POD reduced-order models. Specifically, we normalize the data before POD and use the POD approach to obtain a reduced-order model that accurately approximates the dynamics of the PCCP. Then, we train a multi-layer perceptron (MLP) neural network to capture the dominant dynamics of the POD reduced-order model from the low-dimensional data. The resulting reduced-order POD-based MLP (POD-MLP) model is used as the basis of state estimation. We develop a reduced-order extended Kalman filtering (EKF) algorithm based on the POD-MLP model to estimate the states of the original process. Our approach can effectively reduce the complexity of the NN model and the computational workload required for training while preserving high modeling accuracy. This approach provides a promising solution to state estimation of PCCPs and can potentially be applied to other complex systems with of high dimension for efficient state estimation through bypassing the use of the original higher-dimensional dynamic process model. Our proposed framework offers an advantage in that it can be employed when only data is available. Specifically, in the absence of a physical model, we can still leverage proposed method to build a reduced-order neural network and perform efficient state estimation. 

\section{MODEL DESCRIPTION}\label{Model}

\begin{figure}[t]
	\centering
	\includegraphics[width=\hsize]{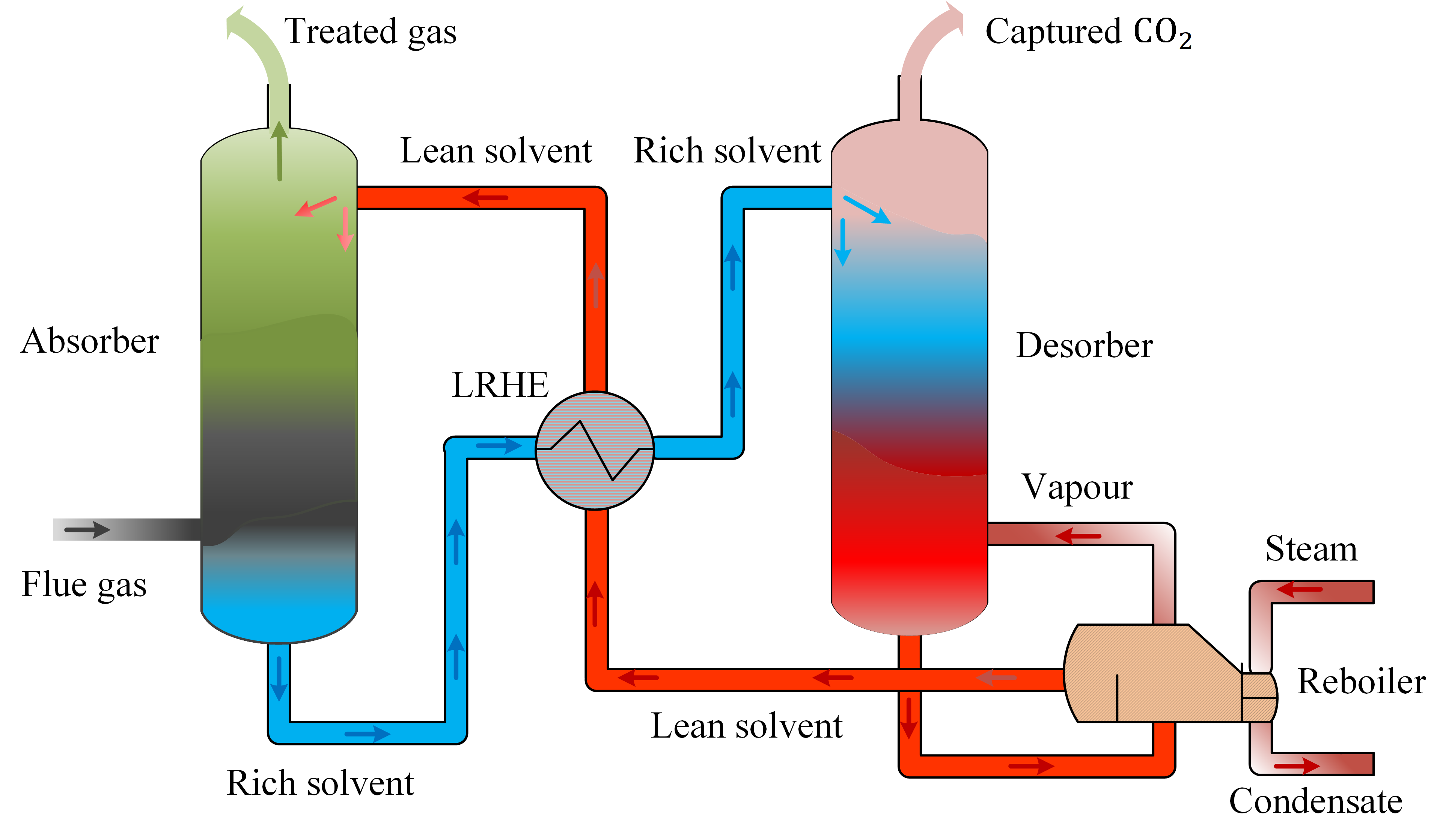}
	\caption{A schematic of the CO$_2$ capture plant.}
	\label{PCCPlant_flowchart}
\end{figure}

Figure~\ref{PCCPlant_flowchart} illustrates the post-combustion capture plant considered in this paper. This diagram shows the four key components of the PCCP, which include the absorption and desorption columns, lean-rich heat exchanger (LRHE), and the reboiler. The flue gas, which contains a high concentration of CO$_2$, is introduced at the bottom of the absorber from the power plant and is then mixed with a lean solvent having low CO$_2$ levels. The 5M Monoethanolamine (MEA) is used as the solvent in this study. The treated flue gas with a reduced amount of CO$_2$ leaves the absorption column, while the rich solvent with a high concentration of CO$_2$ is heated via the heat exchanger by exchanging heat with the lean solvent coming from the reboiler. The rich solvent is then fed into the top of the desorption column, where it is heated through contact with the hot vapor from the reboiler. In the desorption column, the CO$_2$ is stripped from the rich solvent, which is then recycled back to the absorber. The discharged CO$_2$ gas, with a high concentration of CO$_2$ (90-99$\%$), is obtained from the desorption column. 
The details of the PCCP model are briefly described in the following \cite{Decardi-Nelson2018_Process,Harun2012_IJGGC}:
\begin{subequations}\label{lsy12_2a}
\begin{align}
	\label{lsy12_2.1a}
	&\frac{\partial C_L(i)}{\partial t} = \frac{F_{L}}{S_c}\frac{\partial C_L(i)}{\partial l} +(N(i)a^{I}),\\	
	\label{lsy12_2.1b}
	&\frac{\partial C_G(i)}{\partial t} = -\frac{F_{G}}{S_c}\frac{\partial C_G(i)}{\partial l} -(N(i)a^{I}),\\
	\label{lsy12_2.1c}
	&\frac{\partial T_L}{\partial t} = \frac{F_{L}}{S_c}\frac{\partial T_L}{\partial l} +\frac{(Q_La^I)}{\sum_{i=1}^{n}C_L(i)C_{p,i}}, \\
	\label{lsy12_2.1d}
	&\frac{\partial T_G}{\partial t} = -\frac{F_{G}}{S_c}\frac{\partial T_G}{\partial l} +\frac{(Q_Ga^I)}{\sum_{i=1}^{n}C_G(i)C_{p,i}}, \\
	\label{lsy12_2.1e}
	&\frac{dT_{tu}}{dt} = \frac{\dot{V}_{tu}}{V_{tu}}(T_{tu,in}-T_{tu,out}) + \frac{\dot{Q}}{C_{p_{tu}}\rho_{tu}V_{tu}},\\
	\label{lsy12_2.1f}
	&\frac{dT_{sh}}{dt} = \frac{\dot{V}_{sh}}{V_{sh}}(T_{sh,in}-T_{sh,out})+\frac{\dot{Q}}{C_{p_{sh}}\rho_{sh}V_{sh}}, \\
	\label{lsy12_2.1g}
	&\rho C_pV\frac{dT_{reb}}{dt} = f_{in}H_{in}-f_{V}H_{V,out}-f_{L}H_{L,out}+Q_{reb}.
\end{align}	
\end{subequations}
The dynamic models for the absorption column and desorption column are described by the partial differential equations (\ref{lsy12_2.1a})--(\ref{lsy12_2.1d}), where $i=CO_2, MEA, H_2O, N_2$, and the subscripts $L$ and $G$ denote liquid and gas phases, respectively. The dependent variables vary with time $t$ and axial position $l$ of the column. It is assumed that each stage in the two columns is well mixed. Their dynamic models are similar except for a few details like the direction of reactions, temperature, and reaction rate constants. The energy balance equations in (\ref{lsy12_2.1e})--(\ref{lsy12_2.1f}) represent the dynamics of the lean-rich heat exchanger, where $\dot{V}(\rm m^3/s)$ and $\dot{Q}(\rm kJ/s)$ represent the volumetric flow and heat transfer rate, respectively, the subscripts $tu$, $sh$, $in$ and $out$ denote the tube-side, shell-side, inlets and outlets of the heat exchanger, respectively. It is assumed that the mass inside the heat exchanger remains constant. Equation (\ref{lsy12_2.1g}) is the energy balance equation of the reboiler, where $T_{reb}(\rm K)$ represents the temperature of the reboiler, the subscripts $in$, $out$, $V$, and $L$ denote inlet, outlet, vapour and liquid, respectively, and $Q_{reb}(\rm KJ/s)$ is the heat input. The definitions of other variables and parameters of the PCCP can be found in Table~\ref{lsy12_tab_variables}. Physical property calculations of gas and liquid phases are necessary for the model development, and they are estimated from seven nonlinear algebraic correlations. Details of these calculations are not included in this work but can be found in \cite{Decardi-Nelson2018_Process,Harun2012_IJGGC}.

\begin{table} 
	\centering
	\caption{Process variables of each unit of the PCCP.}
	\label{lsy12_tab_variables}
	\renewcommand{\arraystretch}{1.2}
	\tabcolsep 10pt
	\begin{tabular}{lll}\hline
		Notation & Definition & Unit   \\\hline
		$C_i$   & Molar concentrations of component $i$ & $\rm mol/m^3$   \\
		$S_c$   & Cross-sectional area of the column & $\rm m^2$    \\
		$F$     & Volumetric flow      & $\rm m^3/s$   \\
		$N_i$   & Mass transfer rate   & $\rm kmol/m^2s$  \\
		$T$     & Temperature          & $\rm K$  \\
		$l$     & Height of the column & $\rm m$  \\
		$C_p$   & Heat capacity        & $\rm KJ/kmol$ \\
		$Q$     & Heat transfer rate   & $\rm KJ/m^2s$ \\
		$a^{I}$ & Interfacial area     & $\rm m^2/m^3$ \\
		$V$     & Volume               & $\rm m^3$     \\
		$\rho$  & Average molar density& $\rm kmol/m^3$\\
		$H$     & Enthalpy             & $\rm KJ$      \\
		$f$     & Flow rate            & $\rm mol/s$   \\\hline
\end{tabular}\end{table}

The PCCP model consists of partial differential equations for the two columns, and ordinary differential equations for the heat exchanger and reboiler, as well as some algebraic equations for parameter calculations.  As the variables in the columns exhibit temporal and spatial distributions, the partial differential equations are discretized using the method outlined in \cite{Decardi-Nelson2018_Process} to convert them into ordinary differential equations, with the column length divided into five stages. Therefore, the model presented in (\ref{lsy12_2a}) is expressed as a system of differential-algebraic equations (DAEs):
\begin{subequations}\label{lsy12_2.2a}
\begin{align}
	&\bm x(k+1)=\bm F(\bm x(k),\bm a(k),\bm u(k)) + \bm w(k),\\
	&\bm G(\bm x(k),\bm a(k),\bm u(k))=\bm 0,\\
	&\bm y(k)= \bm H(\bm x(k),\bm u(k)) + \bm v(k),
\end{align}
\end{subequations}
where $\bm x(k)\in\R^{103}$ is the state vector, and the definitions of state variables are defined in Table~\ref{lsy11_taba_sensor}, $\bm a(k)\in\R^7$ is the algebraic state vector, $\bm u(k)=[F_L, Q_{reb}, F_G]\in\R^3$ denotes the input vector: solvent flow rate in L/s, reboiler heat in KJ/s, and flue gas flow rate m$^3$/s, and $\bm w(k)$ is the process noise. 

\begin{table}
	\centering
	\caption{Definition of the PCCP state variables at the $j$th discrete point ($j=1,2,\dots,5$).}
	\label{lsy11_taba_sensor}
	\renewcommand{\arraystretch}{1.3}
	\tabcolsep 5pt
	\begin{tabular}{cc|cc|cc}\hline
		States & Def. & States & Def. & States & Def.\\\hline
		$x_{1-5}$   & $C_{L}^j$(N$_2$) & $x_{6-10}$  & $C_{L}^j$(CO$_2$) & $x_{11-15}$ & $C_{L}^j$(MEA) \\\hline
		$x_{16-20}$ & $C_{L}^j$(H$_2$O) & $x_{21-25}$ & $T_{L}^j$ & $x_{26-30}$& $C_{G}^j$(N$_2$) \\\hline
		$x_{31-35}$ & $C_{G}^j$(CO$_2$) & $x_{36-40}$ & $C_{G}^j$(MEA) &
		$x_{41-45}$ & $C_{G}^j$(H$_2$O) \\\hline
		$x_{46-50}$ & $T_{G}^j$ & $x_{51-55}$ & $C_{L}^j$(N$_2$) & $x_{56-60}$ & $C_{L}^j$(CO$_2$) \\\hline
		$x_{61-65}$ & $C_{L}^j$(MEA) & $x_{66-70}$ & $C_{L}^j$(H$_2$O) & $x_{71-75}$ & $T_{L}^j$  \\\hline
		$x_{76-80}$ & $C_{G}^j$(N$_2$) & $x_{81-85}$ & $C_{G}^j$(CO$_2$) & $x_{86-90}$ & $C_{G}^j$(MEA) \\\hline
		$x_{91-95}$ & $C_{G}^j$(H$_2$O) & $x_{96-100}$& $T_{G}^j$ & $x_{101}$ & $T_{h1}$ \\\hline
		$x_{102}$   & $T_{h2}$     & $x_{103}$   & $T_{reb}$ & &\\\hline
	\end{tabular} 
\end{table}
\section{POD AND ITS APPLICATION TO PCCP}

In this section, we employ the proper orthogonal decomposition (POD) to derive a reduced-order model that approximates the dynamics of the PCCP that originally has 103 state variables. By projecting the high-dimensional state variables onto a lower-dimensional subspace, that captures the dominant modes of variability, we obtain a reduced-order model that accurately captures the essential dynamics of the PCCP while significantly reducing its complexity. This is accomplished by computing the singular value decomposition (SVD) of the data matrix, which yields a set of orthonormal basis vectors that represent the most significant patterns of variability in the data.

For general nonlinear systems described by (\ref{lsy12_2.2a}), we obtain a state trajectory by capturing and sampling the system's response to a typical input trajectory at fixed time intervals $\delta$. Then, we sample the resulting state trajectory to construct a matrix of process states from time $0$ to $N$, denoted as:
\begin{align}
\label{lsy12_3a}
	\bfchi=[\bfx(0) \ \bfx(1) \ \dots \ \bfx(N)]\in\R^{n\times(N+1)},	
\end{align}
where the snapshot matrix $\bfchi$ is composed of the actual state at each sampling interval, the number of state variables is denoted as $n$, and the number of sampling intervals is represented by $N$. To ensure a sufficient number of samples, we require $N$ to be much larger than $n$.

For PCCP, the magnitudes of different states vary greatly. To ensure that the POD reduction method is not biased towards states with larger magnitudes in PCCP, each state variable $x_i$ (where $i=1,2,\dots,103$) in the data matrix $\bfchi$ is normalized using (\ref{lsy12_b}) prior to performing SVD decomposition:
\begin{align}
	\label{lsy12_b}
	x_{i,norm} = \frac{x_i-x_{i,min}}{x_{i,max}-x_{i,min}}.
\end{align}
This normalization transforms each state variable $x_i$ to $x_{i,norm}$, where $x_{i,min}$ and $x_{i,max}$ are the minimum and maximum values of $x_i$ in the original dataset, respectively. The resulting normalized matrix $\bfchi_{norm}$ constructed by $x_{i,norm}$ has all states with magnitudes between 0 and 1. SVD is then performed on the normalized matrix $\bfchi_{norm}$ as shown in the following,
\begin{align}
\label{lsy12_3b}
	\bfchi_{norm} = \bfU \bfSigma \bfV^{\T},
\end{align}
where $\bfU\in\R^{n\times n}$ and $\bfV\in\R^{(N+1)\times(N+1)}$ are orthogonal matrices, the rectangular matrix $\bfSigma\in\R^{n\times(N+1)}$ has non-negative real values on its main diagonal. The diagonal entries $\sigma_i$ represent the singular values of the $\bfchi$ matrix, where $i\in{1,2,\dots,n}$. These values are sorted in descending order on the main diagonal of $\bfSigma$.

To construct a reduced-order model, we select a positive integer $r$ that is smaller than the number of states $103$, and truncate $\bfSigma$ at the $r$th column and row to from the reduced-order matrix $\bfSigma_r\in\R^{r\times r}$ using the first $r$ singular values $\sigma_i$. Accordingly, we select the first $r$ columns of $\bfU$ and the first $r$ rows of $\bfV^{\T}$ to form the matrices $\bfU_r$ and $\bfV_r^{\T}$, respectively. Using these matrices, we obtain a reduced-order approximation of normalized process data, given by
\begin{align}
\label{lsy12_3c}
	\bfchi_{norm} \approx \bfU_r \bfSigma_r \bfV_r^{\T}.
\end{align}
We define $\bfxi\in\R^r$ as the state vector of the reduced-order model, and set $\bfxi(k):=\bfU_r^{\T}x(k)$. Using the truncated SVD matrices, the original nonlinear model in (\ref{lsy12_2.2a}) can be expressed as a reduced-order model in state-space form:
\begin{subequations}
\begin{align}
\label{lsy12_3d}
	&\bfxi(k+1) = \bfU_r^{\T}\bfF(\bfU_r\bfxi(k),\bfa(k),\bfu(k))+\bfU_r^{\T}\bfw(k),\\
	&\bfG(\bfU_r\bfxi(k),\bfa(k),\bfu(k))=\bf0,\\
	&\bfy(k) = \bfH(\bfU_r \bfxi(k),\bfu(k))+\bfv(k).
\end{align}
\end{subequations}
The evolution of $\bfxi(k)$ in the reduced-order model can be used to approximate the actual state trajectory of the original nonlinear process through the mapping $\bfx(k)\approx \bfU_r\bfxi(k)$.
\begin{remark}
	In this section, we improve the accuracy of the model approximation by normalizing the data prior to applying the POD method to obtain a reduced-order model. The effectiveness of the normalization for POD will also be illustrated through simulations in Section~\ref{Examples}.
\end{remark}

\section{APPROXIMATING REDUCED-ORDER MODEL WITH MLP NETWORKS}

The POD technique is commonly applied in linear systems to decrease computational costs by reducing the dimensionality of the problem. However, it does not yield similar advantages for nonlinear systems due to the challenge in explicitly expressing $\bfU_r^{\T}\bfF(\bfU_r\bfxi,\bfa,\bfu)$ in terms of the reduced basis $\bfU_r$. Consequently, evaluating the reduced-order model may require more time than evaluating the original nonlinear function $\bfF$. Different approaches have been adopted to address this issue. For instance, a linear parameter varying model was utilized to approximate the reduced-order model, however, the benefits were found to be insignificant \cite{Zhang2019_Process}.

To address this issue, we present a method to speed up the evolution of reduced-order models for nonlinear systems such as (\ref{lsy12_2.2a}). The method employs a multi-layer perceptron (MLP) neural network to fit the reduced-order model and hence decrease the computation time. The MLP neural network model of $\bfxi$ is given by:
\begin{align}
\label{lsy12_mlp}
    \bfxi(k+1) = \bff_{mlp}(\bfxi(k),\bfu(k)) + \bfw_r(k),
\end{align}
where the vector function $\bff_{mlp}\in\R^r$ approximates the dynamic behavior of $\bfxi$ in (\ref{lsy12_3d}), and $\bfw_r\in\R^r$ denotes the process noise and model error. Consequently, we do not have to evaluate the vector function $\bfF: \R^{103} \rightarrow \R^{103}$ of the full-order model, leading to significant time savings.

MLP models are a type of artificial neural network widely used to approximate complex and nonlinear problems. An MLP typically consist of an input layer, one or more hidden layers, and an output layer. Each layer in an MLP contains multiple fully connected neurons, which are connected to the next layer by weights. In supervised learning, the weights are adjusted to approximate each target value. The number of neurons in the input and output layers is determined by the input and output variables, respectively. The computation time required for MLP output is relatively short, as only a few matrix multiplications, vector additions, and function evaluations are necessary. Given these characteristics, MLP is a suitable choice for approximating the vector function $\bff_{mlp}$ in equation (\ref{lsy12_2.2a}) after POD model reduction. The basic model formulation of MLP is indicated below:
\begin{align}
\label{lsy12_4a}
	\bfz^{(l)}=\sigma_h(\bfw^{(l)} \bfz^{(l-1)}+\bfb^{(l)}), \quad \bfy=\bfw^o \bfz+\bfb^o
\end{align}
where $\bfz^{(l)}$ denotes the output vector of the $l$-th hidden layer, obtained by applying an activation function $\sigma_h$ to the weighted sum of the input vector $\bfz^{(l-1)}$ and the bias vector $\boldsymbol{b}^{(l)}$. The weight matrix of the $l$-th hidden layer is denoted by $\bfw^{(l)}$. The input vector $\bfz^{(0)}$ is the MLP input, and the output vector $\bfy$ is obtained by applying the weight matrix $\bfw^o$ to the output vector of the last hidden layer and adding the bias vector $\bfb^o$. It is worth noting that the MLP output layer is typically linear, while the activation function of the hidden layer can be chosen from various options based on the specific problem being solved.

\begin{figure}[t]
	\centering
	\includegraphics[width=\hsize]{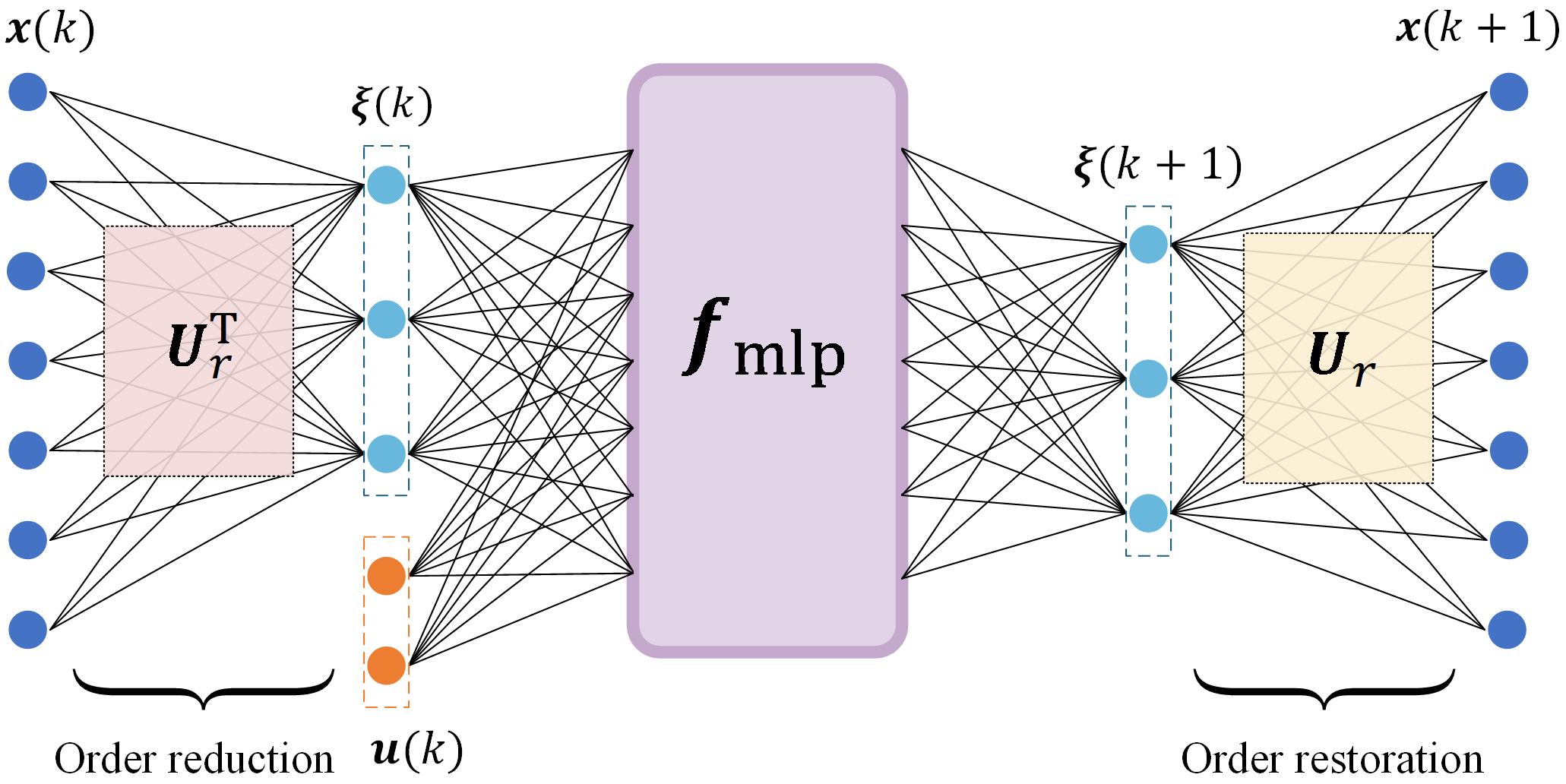}
	\caption{The diagram of the POD-MLP model.}
	\label{lsy12_fig:mlp_model}
\end{figure}

The reduced-order PCCP model is enhanced by utilizing an MLP network, where the input is $\bfxi_u:=[\bfu^{\T}, \bfxi^{\T}]^{\T}\in\R^{3+r}$, and the output is $\hat{\bfxi}\in\R^r$. The input layer has $3+r$ neurons, and the output layer has $r$ neurons. The MLP model is trained to minimize the mean-squared-error (MSE) between the predicted output $\hat{\bfxi}$ and the actual output $\bfxi$:
\begin{align}
	L = MSE(\bfxi,\hat{\bfxi}).
\end{align}
The model structure of the proposed POD-based MLP (POD-MLP) model is shown in Figure~\ref{lsy12_fig:mlp_model}.

\begin{remark}
	As the system dimension increases, the number of training samples required to construct an MLP model that accurately approximates the system grows exponentially. Therefore, direct training a large-scale nonlinear model can be computationally expensive. Hence, performing POD dimension reduction for high-dimensional PCCP models and then training the MLP model for the reduced-order model is a highly practical approach.
\end{remark}

\begin{remark}
	The POD-MLP model not only reduces the order of the system, but also eliminates the need for solving the DAEs in the original PCCP model. It achieves this by extracting the dynamic information of the algebraic state variable $\bfa$ from the data, and excluding $\bfa$ from the model.
\end{remark}

\section{EKF USING the POD-MLP MODEL}

In this section, we develop an extended Kalman filtering (EKF) based on the POD-MLP model (the POD-MLP-EKF algorithm for short) to estimate the actual process states. The POD-MLP model is summarized in the following:
\begin{subequations}
\begin{align}
	&\bfxi(k+1) = \bff_{mlp}(\bfxi(k),\bfu(k)) + \bfw_r(k),\\
	&\bfy(k) = \bfH(\bfU_r \bfxi(k),\bfu(k))+\bfv(k).
\end{align}
\end{subequations}
Assuming that $\bfw_r(k)$ and $\bfv(k)$ are two mutually uncorrelated Gaussian noise sequences with zero-mean, we further assume that they have covariance matrices $\bfQ_r$ and $\bfR_r$.

Based on the above preparation, the POD-MLP-EKF algorithm is designed in the following two steps:\\
\noindent{Step 1: Prediction step:}
\begin{subequations}
\begin{align}
	&\hat{\bfxi}(k+1|k) = \bff_{mlp}(\hat{\bfxi}(k|k),\bfu(k)),\\
	&\bfP(k+1|k) = \bfA(k)\bfP(k|k)\bfA^{\T}(k) + \bfQ_r,
\end{align}
\end{subequations}
where $\hat{\bfxi}(k+1|k)$ is the prediction of the system state at time $k+1$ based on the current state estimate $\hat{\bfxi}(k|k)$ and the input $\bfu(k)$, and $\bfP(k+1|k)$ contains a priori error covariance information, incorporating the prediction error and the uncertainty associated with the system dynamics through the covariance matrix $\bfQ_r$. The matrix $\bfA(k)=\frac{\partial\bff_{mlp}(\bfxi,\bfu)}{\partial\bfxi}|_{\bfxi=\hat{\bfxi}(k|k)}$ is the Jacobian matrix of the MLP model with respect to the state vector $\bfxi$ evaluated at the predicted state estimate $\hat{\bfxi}(k|k)$.\\
\noindent{Step 2: Update step:}
\begin{subequations}
\begin{align}
	&\bfK(k+1)=\frac{\bfP(k+1|k)\bfC^{\T}(k)}{\bfC(k)\bfP(k+1|k)\bfC^{\T}(k)+\bfR_r},\\
	&\hat{\bfxi}(k+1|k+1)=\hat{\bfxi}(k+1|k)+\bfK(k+1)[y(k+1)\non\\
	&\quad\quad\quad\quad\quad\quad\quad \ -\bfH(\bfU_r \hat{\bfxi}(k+1|k),\bfu(k))],\\
	&\bfP(k+1|k+1)=[\bfI-\bfK(k+1)\bfC(k)]\bfP(k+1|k),
\end{align}
\end{subequations}
where the correction gain $\bfK(k+1)$ is computed based on the a priori estimation error covariance $\bfP(k+1|k)$, the measurement error covariance $\bfR_r$, and the observation matrix $\bfC(k)$, which maps the predicted state $\hat{\bfxi}(k+1|k)$ to the measurement space. The state estimate is then updated to $\hat{\bfxi}(k+1|k+1)$ using the correction gain and the measurement innovation $y(k+1)-\bfH(\bfU_r\hat{\bfxi}(k+1|k),\bfu(k))$, which represents the difference between the actual measurement and the predicted measurement based on the predicted state. Finally, the a posteriori estimation error covariance matrix $\bfP(k+1|k+1)$ is computed based on the updated state estimate and the correction gain, which reflects the reduced uncertainty in the estimated state after incorporating the measurement information.

Then, we can obtain the state estimates of the actual PCCP states, denoted by $\hat{\bfx}$, by utilizing the reduced-order state estimate $\hat{\bfxi}(k+1|k+1)$ and the linear mapping $\bfU_r$, as follows:
\begin{align}
	\hat{\bfx}(k+1) = \bfU_r \hat{\bfxi}(k+1|k+1).
\end{align}

\section{SIMULATION RESULTS}
\label{Examples}

\begin{figure}[t]
	\centering
	\includegraphics[width=\hsize]{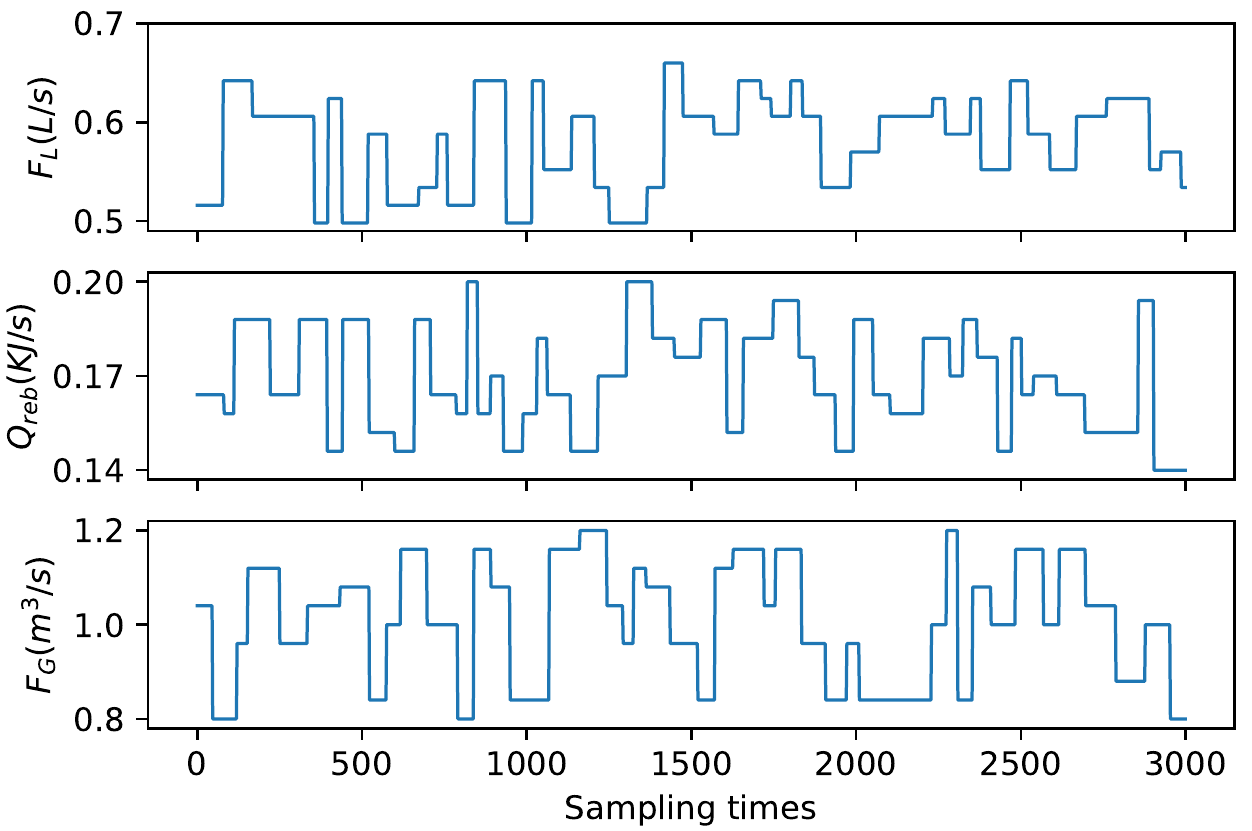}
	\caption{The three input signals for generating system state trajectories in model reduction.}
	\label{lsy12_fig_input}
\end{figure}

To perform the model order reduction, the input vector $\bfu(k)=[F_L, Q_{reb}, F_G]$ is used to excite the PCCP, which are constrained as 0.48 L/s $\leq F_L \leq$ 0.66 L/s, 0.14 KJ/s $\leq Q_{reb} \leq$ 0.20 KJ/s and 0.8 m$^3$/s $\leq F_G \leq$ 1.2 m$^3$/s. The dynamic model of the PCCP is
discretized at a sample interval of $\Delta = 30$s. Pseudo-random multi-level signals (PRMSs) are commonly used as excitation signals for identifying nonlinear systems. For example, a ten-level PRMS used in the PCCP is shown in Figure~\ref{lsy12_fig_input} (Samples 1 to 3000). The switching times are randomly chosen from a uniform distribution ranging between 900s and 3000s (30 to 100 sampling times). The system states are sampled over a duration of 100 hours to construct the snapshot matrix $\bfchi$ for POD reduction. With 12,000 sampling intervals, the requirement $N\gg n$ is satisfied.

\begin{figure}[t]
	\centering
	\includegraphics[width=0.9\hsize]{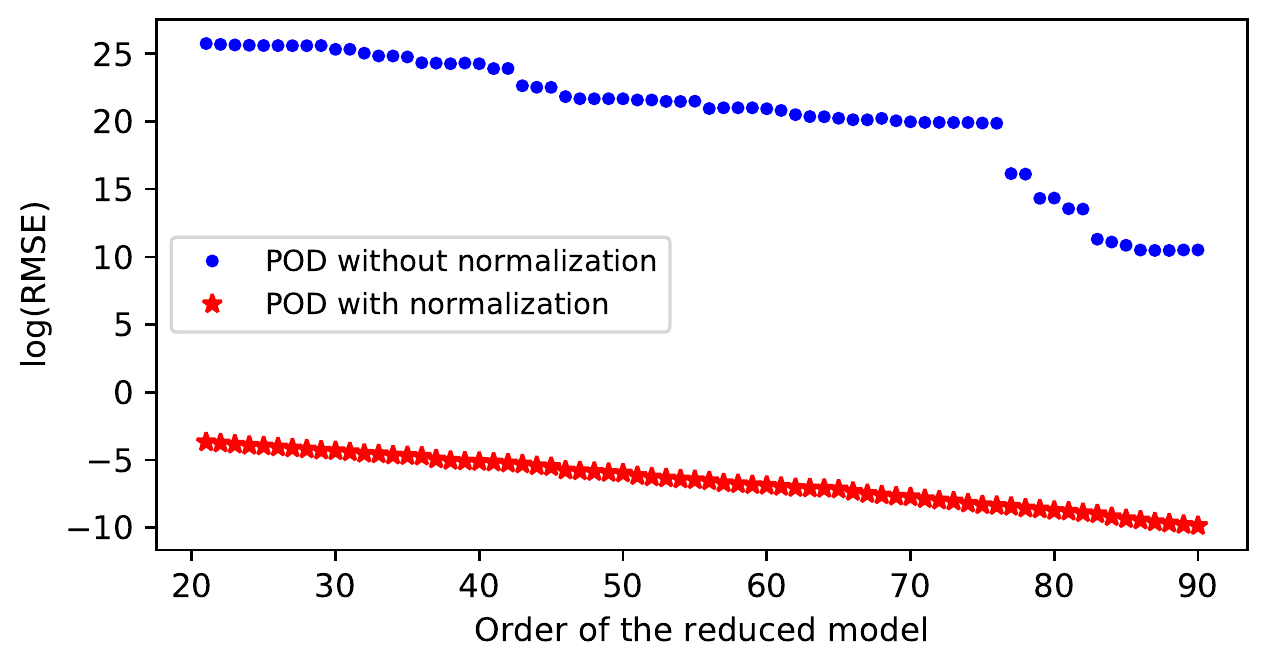}
	\caption{The values of $log$(RMSE) at different orders.}
	\label{lsy12_fig_RMSE}
\end{figure}

\begin{figure}[t]
	\centering
	\includegraphics[width=\hsize]{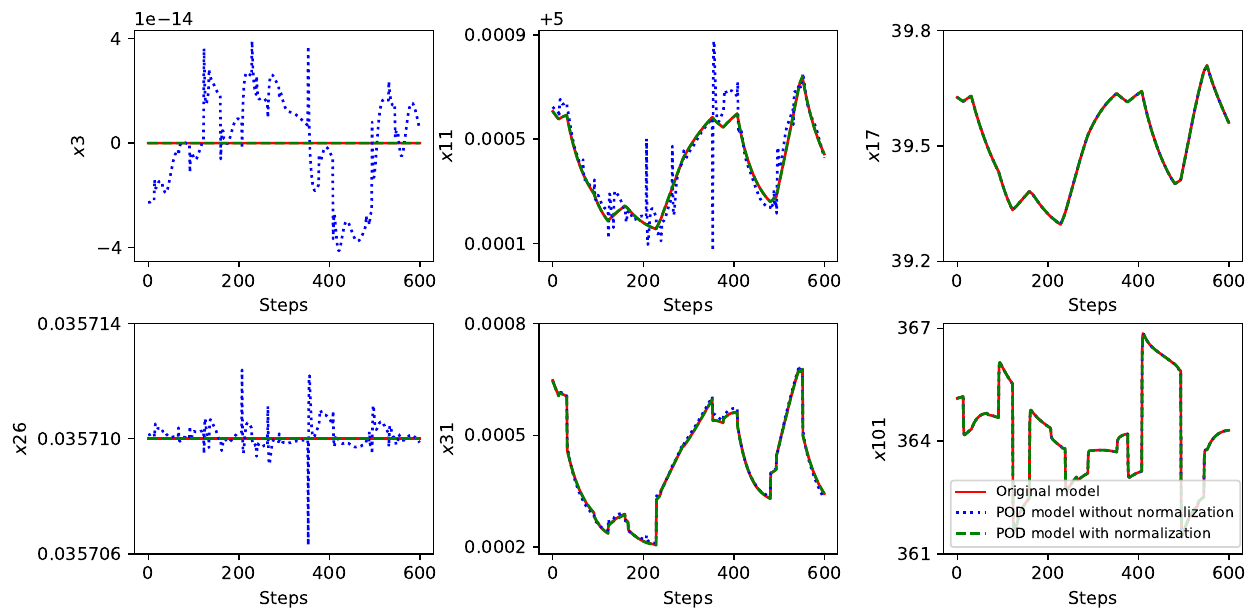}
	\caption{The trajectories of some of state based on the original model and the reduced model using POD with normalization and POD without normalization.}
	\label{lsy12_fig_2}
\end{figure}

Next, we apply the SVD to the matrix $\bfchi$ which has dimensions 103 by 12000. This results in a unitary matrix $U$ that can be used for coordinate transformation. To evaluate the reduced-order model accuracy, we use the root-mean-square error (RMSE) to denote the model errors, which is defined as follows:
\begin{align}
	{\rm RMSE}&:=\sqrt{\frac{\sum^{N}_{j=0}\sum^{103}_{i=1}(x_{i,norm}(j)-\hat{x}_{i,norm}(j))^2}{N}},\non
\end{align}
where $\hat{x}_{i,norm}:=\bfU_r \xi_i$ is the $i$th approximated state obtained from a reduced order model. To validate the accuracy of a reduced-order model, we should use input trajectories that are different from the ones used in POD model reduction. Additional 600 sampling data are used for validation. Based on the actual state trajectories and reduced-order model state trajectories, the RMSE is calculated for each model. The values of log(RMSE) at different orders $r=20, \dots, 90$ under the POD with normalization and the POD without normalization are shown in Figure~\ref{lsy12_fig_RMSE}. It shows that the degree of model mismatch increases with the decrease in the model order for both cases. Moreover, the RMSE value of POD with normalization is smaller than that without normalization. This is because for the POD method without normalization of the data matrix $\bfchi$, the approximation of the state with small numerical values is not accurate when the model is reduced, and the model approximation error is added to each state in the form of absolute value. While the method that normalizes the data matrix $\bfchi$ reflects the model approximation error in the form of relative values for each state. Therefore, RMSE values of reduced-order models at different orders obtained from POD with normalization are smaller than those obtained from POD. This can also be further demonstrated from Fig.~\ref{lsy12_fig_2}, which shows the trajectories of some of states based on the original model and the reduced model with order 30 using POD with normalization and POD methods respectively. From the figure, it can be seen that for states with large numerical values ($x_{17}$, $x_{101}$), both methods provide very good approximation effects, but for states with small numerical values or small fluctuations ($x_{3}$, $x_{11}$, $x_{26}$, $x_{31}$), the approximation of POD with normalization is significantly better than that of POD.

\begin{figure}[t]
	\centering
	\includegraphics[width=\hsize]{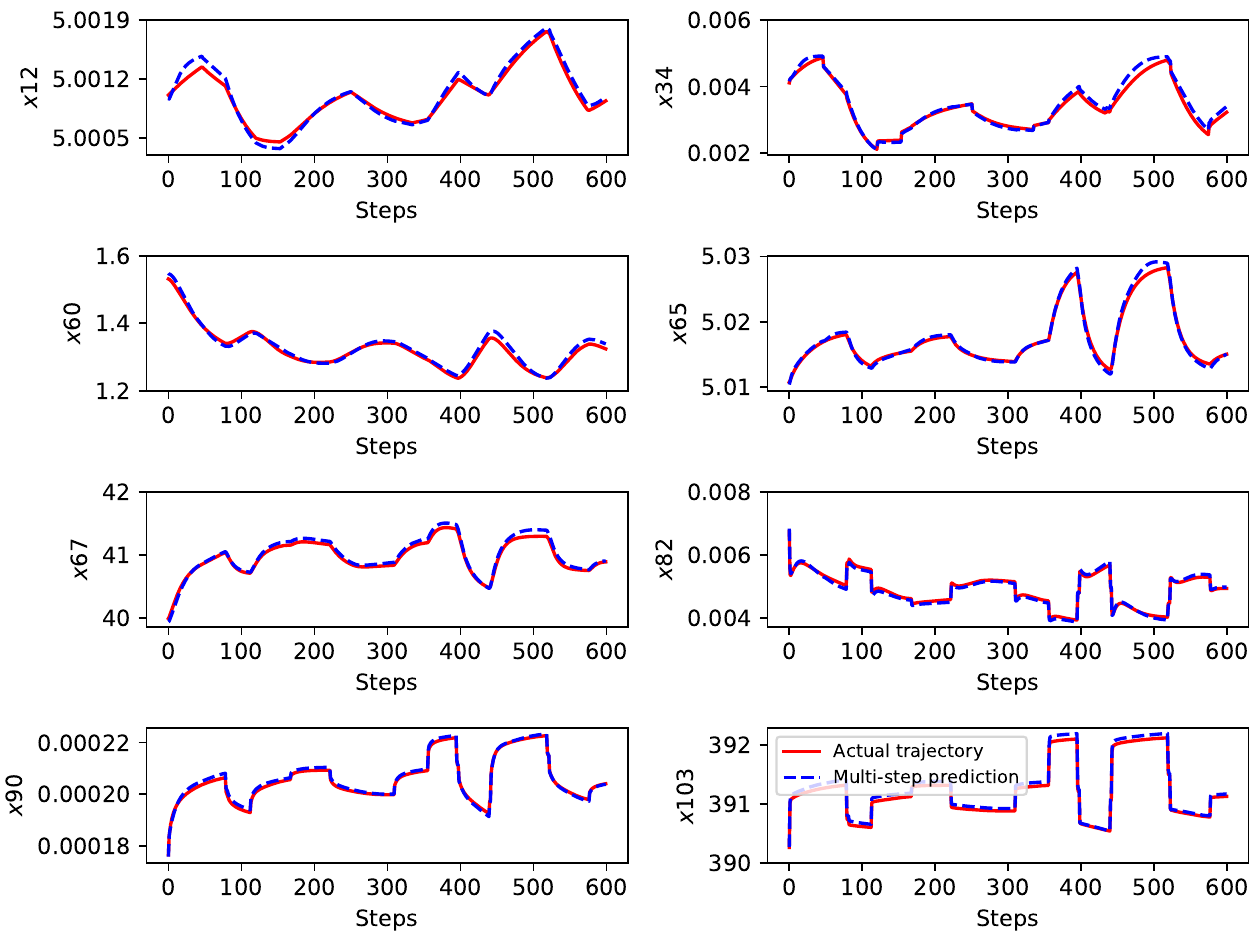}
	\caption{Multi-step prediction for the PCCP.}
	\label{lsy12_fig_3}
\end{figure}

Next, we will evaluate the open-loop prediction performance of the POD-MLP model. Using the PRMS input signal, we generate 100,000 samples for each state by utilizing the first-principle model with $\bfU_r$. These data are split into training (70$\%$), validation (20$\%$), and testing (10$\%$) datasets. The developed POD-MLP model consists of 3 hidden layers with 128 neurons, the input of the POD-MLP is the normalized 33-dimensional vector $\bfxi_u$ and the output is the normalized state vector $\bfxi$. The activation function for the hidden layers has been selected to be Tanh, and a linear activation function is used in the output layer. Figure~\ref{lsy12_fig_3} demonstrates the testing performance of the POD-MLP model in multi-step open-loop prediction for the actual state trajectory of the PCCP. It can be seen that the fit is accurate.

\begin{figure}[t]
	\centering
	\includegraphics[width=\hsize]{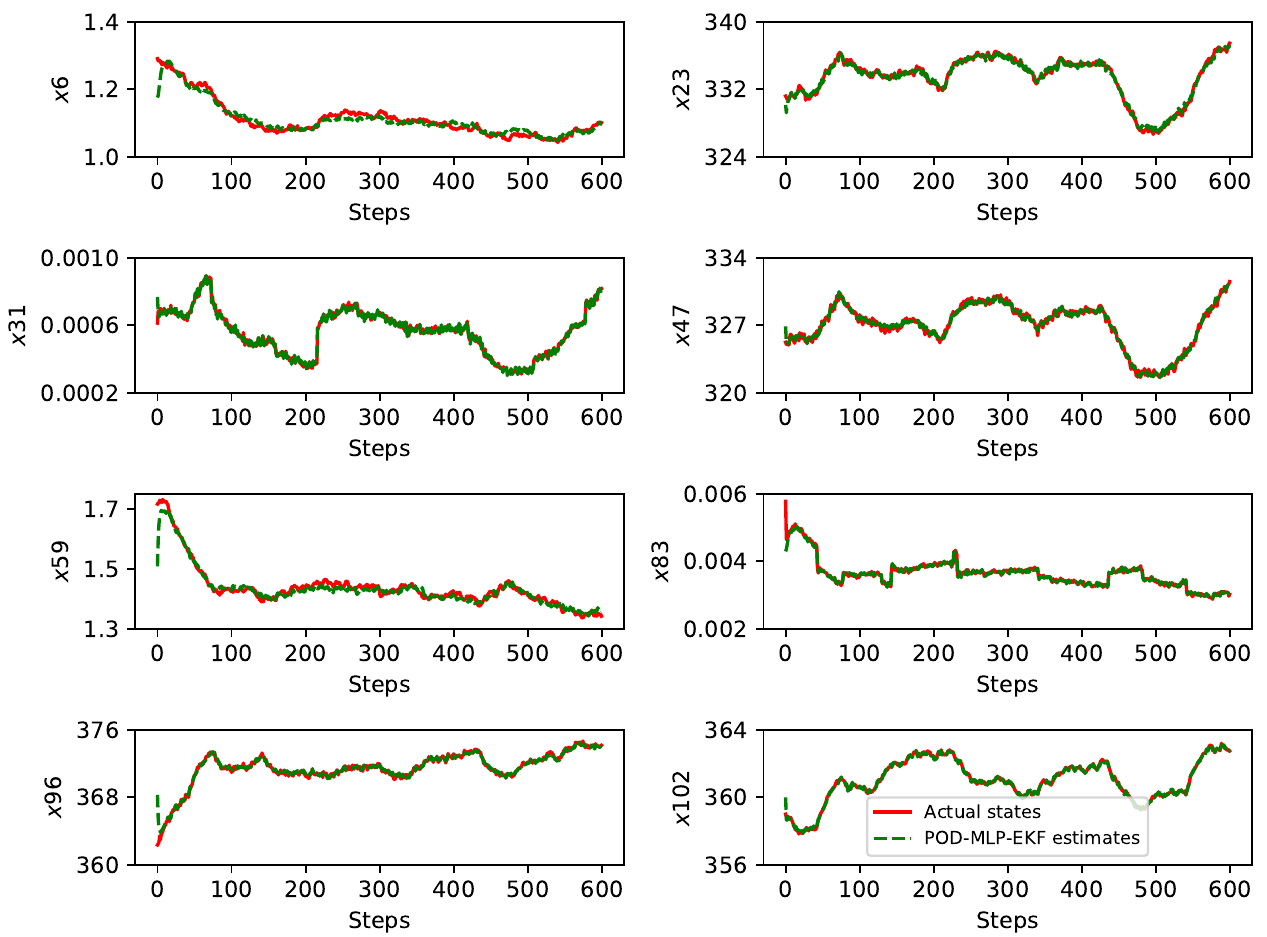}
	\caption{The actual states and POD-MLP-EKF state estimates for the PCCP.}
	\label{lsy12_fig_4}
\end{figure}

\begin{table}[t] 
	\centering
	\caption{Running time.}
	\label{lsy12_tab}
	\begin{tabular}{|c|c|c|c|}\hline
		& EKF & POD-EKF & POD-MLP-EKF \\\hline
		Model prediction    & 8s   & 8s   & 3s  \\\hline
		Discretization      & 115s & 117s & 3s  \\\hline
		Other steps for EKF & 2s   & 2s   & 0.4s \\\hline
		Total time (s)      & 125s & 127s & 6.4s \\\hline
	\end{tabular}
\end{table}
  
In the following analysis, we focus on the state estimation of the PCCP using the POD-MLP-EKF algorithm. The PCCP is subject to process disturbances, and the output measurements are corrupted by random noise. Specifically, each process disturbance sequence associated with the $i$th state $x_i$ is generated following normal distribution with zero mean and a standard deviation 0.01$x_{i,s}$, where $x_{i,s}$ is the steady-state value of $x_i$. Random noise is added to each measurement $y_i$ as Gaussian white noise with zero mean and a standard deviation 0.01$y_{i,s}$, where $y_{i,s}$ is the value of $y_i$ at steady-state. As a result, the covariance matrices of process noise and measurement noise are $\bfQ={\rm diag}((0.01\bfx_s)^2)$ and $\bfR={\rm diag}((0.01\bfy_s)^2)$, respectively. The tuning parameters in the POD-MLP-EKF are $\bfQ_r=\bfU_r^{\T} \bfQ \bfU_r$ and $\bfR_r=\bfU_r^{\T} \bfR \bfU_r$. The initial guess for the normalized states is set to $0.5\bf1_{103}$. Figure~\ref{lsy12_fig_4} shows some of the state estimates and the actual states. The proposed estimation scheme provides accurate state estimates.

We compare the computational efficiency of the proposed POD-MLP-EKF algorithm based on model reduction and neural network, a centralized EKF design directly based on the PCCP model, and an EKF for POD model, denoted as the POD-EKF. Specifically, we evaluate the average computation time for 600 steps required by the three algorithms, which is 125 s, 127 s, and 6.4 s, respectively. The results demonstrate that the combination of POD reduction and neural network significantly reduces the computing time.

\addtolength{\textheight}{-12cm}   



%

\section*{ACKNOWLEDGMENT}
The first author, S.Y. Liu, is a visiting Ph.D. student in the Department of Chemical and Materials Engineering at the University of Alberta from March 2021 to February 2023. She acknowledges the financial support from the China Scholarship Council (CSC) during this period.



\end{document}